\documentstyle[a4]{article}
\begin{document}

\newcommand{\be}{\begin{equation}}
\newcommand{\ee}{\end{equation}}
\newcommand{\epe}{\end{equation}}
\newcommand{\bea}{\begin{eqnarray}}
\newcommand{\eea}{\end{eqnarray}}
\newcommand{\ba}{\begin{eqnarray*}}
\newcommand{\ea}{\end{eqnarray*}}
\newcommand{\epa}{\end{eqnarray*}}
\newcommand{\ar}{\rightarrow}

\def\s{\sigma}
\def\r{\rho}
\def\D{\Delta}
\def\R{I\!\!R}
\def\l{\lambda}
\def\g{\gamma}
\def\D{\Delta}
\def\cD{{\cal D}}
\def\cH{{\cal H}}
\def\dA{A^{\dag}}
\def\d{\delta}
\def\T{\tilde{t}}
\def\X{\tilde{X}}
\def\ts{\tilde{\sigma}}
\def\k{\kappa}
\def\t{\tau}
\def\f{\phi}
\def\p{\psi}
\def\z{\zeta}
\def\ep{\epsilon}
\def\hx{\widehat{\xi}}
\def\A{\tilde{a}}
\def\B{\tilde{b}}
\def\a{\alpha}
\def\b{\beta}
\def\O{\Omega}
\def\H{\cal H}
\def\tH{\tilde{H}}
\def\M{\cal M}
\def\g{\hat g}
\newcommand{\dslash}{\partial\!\!\!/}
\newcommand{\aslash}{a\!\!\!/}
\newcommand{\eslash}{e\!\!\!/}
\newcommand{\bslash}{b\!\!\!/}
\newcommand{\vslash}{v\!\!\!/}
\newcommand{\rslash}{r\!\!\!/}
\newcommand{\cslash}{c\!\!\!/}
\newcommand{\fslash}{f\!\!\!/}
\newcommand{\Dslash}{D\!\!\!\!/}
\newcommand{\Aslash}{{\cal A}\!\!\!\!/}


\begin{center}

{\large D-branes as coherent states in the open string
channel.}

\vspace*{1.2cm} {\large M. Botta Cantcheff $^{\ddag}$
\footnote{e-mail: botta@cbpf.br, botta@fisica.unlp.edu.ar}}

\vspace{3mm}

$^{\ddag}$ Instituto de Fisica La Plata, CONICET, UNLP \\
CC 67, Calles 49 y 115, 1900 La Plata, Buenos Aires, Argentina

\end{center}

\begin{abstract}
\noindent

We show that bosonic D-brane states may be represented as coherent states in an open string representation.
 By using the Thermo Field Dynamics (TFD) formalism,
 we may construct a condensed state of open string modes which encodes the information on the D-brane configuration.

 We also introduce a construction alternative to TFD, which does not requires to assume thermal equilibrium.
It is shown that the dynamics of the system combined with geometric
properties of the duplication rules of TFD is sufficient to obtain
the thermal states and their analytic continuations in a geometric
fashion. We use this approach to show that bosonic D-brane state in
the open string sector may also be built as boundary states in a
special sense.

 Some implications of this study on the interpretation of the open/closed duality
  and on the kinemathic/algebraic structure of an open string field theory are also commented.

\end{abstract}

\section{Introduction}
\

  D-branes states may be constructed as boundary states
 in the closed string Hilbert space by using
the so-called world sheet duality \cite{boundarystates}, however since
one may define D-branes as the
surfaces where the \emph{open strings} end, 
it is natural to ask for this definition also in the open string description.
In this sense, it has been shown that some exact classical solutions of Vacuum string field theory, which is a simplification of the Witten's open string field theory, represent D-branes \cite{RSZ}. 
Other approaches addressed to define boundary states in the open channel where recently proposed \cite{matsuo}.
Apart from this,
the possibility of describing D-branes in terms of open string states,
 in contrast to the traditional approaches (in the closed string channel) is specially motivating since
  it might shed some light on the open/closed duality, intimately related to the AdS/CFT conjecture.

On the other hand,
 by virtue of the microscopical description of the black hole entropy \cite{bh},
one of the most interesting problems concerning D-branes is the development of a model where their thermodynamical properties and microscopical structure be clarified. The framework presented in this article is devoted to the two purposes simultaneously.

 The problem with
defining D-brane states in the open string channel is that the boundary
conditions are imposed on the operators of the theory rather on
particular states,
 which could be
interpreted as D-brane states in an open string Fock space. However,
  we claim that this difficult may be solved
in the context of the so called Thermo Field Dynamics (TFD), developed by Takahashi
and Umezawa \cite{ume2,ume4,rev2,ume1,kha2,kha3}, where
an identical but fictitious copy of the system is introduced. In
this framework, variables and degrees of freedom are duplicated so
as the original Hilbert
  space state, then it seems to be possible to construct a state (or a family of states) to describe an open-string + brane system.

Thermo Field Dynamics 
is a real time approach to
quantum field theory at finite temperature \cite{kob1,leb1}. In this formalism one canonically quantizes the
 fields as operators on a \emph{thermal} Hilbert space and the statistical average of an operator
 $Q$ is defined as its expectation value
in a thermal vacuum state:
\begin{equation}\frac{{\mbox Tr}[Q e^{-\b H}]}{Z}
= \left\langle 0 (\b ) \left| Q
\right| 0 (\b )\right\rangle .
\label{tr1}
\end{equation}
The hamiltonian evolution of these thermal fields is given by the
operator $H-\tilde{H}$, where $H$ and $\tilde{H}$ denote the
hamiltonian of the original system and its non-physical copy
respectively. So the fundamental state encoding the statistical
information may be represented as follows \be |0(\b) \rangle
\rangle= Z^{-1 /2} \sum_n  e^{ -\b E_n /2 } | n \rangle
|\tilde{n} \rangle , \ee where $| n ; \tilde{n} \rangle$ denotes the
$n^{th}$ energy eigenvalue of the two systems.

The idea of using TFD to study D-branes at finite temperature
 came up in Refs. \cite{IVV,AGV2,AEG,AGV3,AGV4,AGV5},
 where thermal boundary states are constructed in the closed string channel by considering string coordinates as
 thermal
  fields. In contrast, one may compute the free energy of the open string and obtain
   the self-energy of the D-brane by using the finite temperature dualities \cite{vazquez}. However this method does
    not provide us the representation of the D-brane as a (thermal) state in the open string channel.
 The general goal of this work is precisely obtain such D-brane states at arbitrary finite temperature in general.

The paper is organized as follows. In Section 2 we briefly describe the TFD formalism and propose 
that a Dp-brane states may be built as the fundamental one in a thermal Hilbert space of open strings. We show this statement using the standard 
 representation of boundary states in Subsection 2.1.
In Section 3 we develop a geometric approach in order to formulate D-branes as boundary states in the
 open string channel. Finally, in Section 4 we present the conclusions of this approach and discuss the main consequences and perspectives.

\section{Thermo Field Dynamics and D-brane states}

Let us consider the thermodynamics of an open string, that is, to
consider a bosonic open string in contact with a thermal reservoir
at temperature $\b^{-1}$. The partition function in the canonical
ensemble is \be\label{path} Z_o(\b ) = tr \, e^{-\b \, H_o} =
\int D X e^{-S_{W_o} [X]} \ee where the Euclidean,
two-dimensional world sheet manifold is $W_o \sim S^{1}_{\b}\times
[0, \pi]$ ($\b$ denotes the circle length)
 where the path integral is realized by
summing over histories of the open string fields $X: W_o \to M$
satisfying specific boundary conditions which encodes the
information about the Dp-brane. Since $\partial W_o =S^{1}_{\b, -} \cup
S^{1}_{\b, +}$
 where the
circles $S^{1}_{\b,\pm}$ correspond to the points $0,\pi$
respectively and the corresponding boundary conditions are:

\be\label{dir}
 X^i|_{\s=0, \pi}= x^i_{\pm } \;\;\; i=p+1,...25
\ee \be\label{neu}
\partial_\s X^a|_{\s=0,\pi }=0 \;\;\; a=0,1,..., p
\ee

For simplicity, we assume here that both string endpoints belong to the same
Dp-brane.
If one assumes that an open string interchanges energy-momentum only through its endpoints, then the reservoir
 should thought to be placed in the same region that these two point are confined. This is precisely the
 Dp-brane surface. So a priori, one could be tempted to identify the Dp-brane with the macroscopical system
 which acts as the thermal reservoir in itself.

Let us observe that the density matrix $\rho =
Z^{-1}e^{-\b \, H_o} $ describes the thermal (mixed) state the
bosonic open string attached to the Dp-brane in equilibrium, and it
encodes the full information about this
 system in the open channel.
Therefore, by virtue of the TFD approach, the thermal state given by this
density matrix $\r $ is \emph{equivalent} to a pure state (the thermal vacuum) in the tensor product of two copies of the quantum Hilbert space of open string, which encloses the quantum and thermal information of the D-brane. 
In this way, we have a straightforward procedure to construct a representation of a D-brane as a well defined coherent state directly in the open channel and we do not need to use the standard world-sheet transformation \cite{boundarystates} to construct the boundary states in the closed picture. For
completeness however, in Subsection 2.1, we will show that in fact this thermal vacuum corresponds to a boundary state via the world-sheet correspondence.

\vspace{0.6cm}

As mentioned before, the TFD algorithm consists first in duplicating
the degrees of freedom of the system.
To this end a copy of the original
Hilbert space may be constructed with a set of operators of creation/anihilation
that have the same commutation properties as the original
ones. The total Hilbert space is the tensor product of the
two spaces ${\cal H}_{o}\otimes\widetilde{{\cal H}}_{o}$, where in this case ${\cal H} _{o}$
 denotes the physical quantum states space of the bosonic open string whose endpoints are in contact with Dp-brane
described by (\ref{dir} , \ref{neu}).

From now on let us adopt the light cone gauge and use the capital indices $I;J;K...=1,...,D-1$ to denote the
 physical components ($\mu= I, +,-$) of the string embedding.
 The general solution, for the open string coordinates  satisfying these boundary conditions
 reads \be\label{stringsoldir}  X^i (t,\sigma)= x^i_{-}
\;(\pi-\s)/\pi + x^i_{+ } \; (\s/\pi) - \sqrt{2\a'}
\sum_{n \neq 0}\; \left(\frac{1}{n}\;\a_n^i \;e^{-int}\; \sin(n\s)\right) \ee

\be\label{stringsolneu} X^a(t,\sigma)= x^a + 2\a' p^a t + 2i \a'
\sum_{n \neq 0} \left( \frac{1}{n} \;\a_n^a \; e^{-int} \;\cos(n\s)\right) \ee
  The solution correspondent to the non-physical string $\X({\tilde t},{\tilde \s})$, with the
   same boundary conditions, may also be expanded in this basis of solutions.  Since introducing finite
    temperature breaks the Lorentz invariance, we consider these solution in the zero-momentum frame: $p^a={\tilde p^a }=0$.
 Next, these fields are canonically quantized and, according to the TFD rules \cite{kha5} the operators of the
  two system are built commuting among
themselves, so the doubled system is described by two
independent strings defining two world-sheets.

The Fourier modes may be redefined as,
\begin{eqnarray}\label{redef}
a_n^I =  \frac{\a_n^I}{\sqrt{n}}\nonumber
\\
a_n^{\dagger \:I} = \frac{\a_{-n}^I}{\sqrt{n}} \, ,
\end{eqnarray}
so as the tilde oscillators, in order to
satisfy the extended algebra:
\begin{eqnarray}
\left[a_{n}^I,a_{m}^{\dagger \:J}\right]
&=&\left[\tilde{a}^{I}_{n},\tilde{a}_{m}^{\dagger\:J}\right]
=  \delta_{n,m}\delta^{I,J},\label{alg}
\nonumber
\\
\left[a_{n}^{\dagger\:I},\tilde{a}_{m}^{J}\right]
&=&\left[a_{n}^{\dagger\:I},\tilde{a}_{m}^{\dagger\:J}\right]
=\left[a_{n}^{I},\tilde{a}_{m}^{J}\right]=
\left[a_{n}^{I},\tilde{a}_{m}^{\dagger\:J}\right]=0.
\end{eqnarray}
 The standard vacuum in this extended theory is defined by
\be {a}^{I}_{n}\left.\left|0\right\rangle \! \right\rangle= \A_{n}^I
\left.\left|0\right\rangle \! \right\rangle = 0
  , \ee for $n>0$ and $\left.\left|0\right\rangle \!
\right\rangle =|0\rangle \otimes|\widetilde{0} \rangle$ as usual.
However, the physical thermal fundamental state shall be obtained
from this through a Bogoliubov transformation, $e^{-i{G}}$, which
entangles the states of the two independent Hilbert spaces. This is
given by the following relation \be \left |0(\theta)\right\rangle =
e^{-i{G}} \left.\left|0\right\rangle \! \right\rangle =
\prod_{n=1}\left[\left( \frac{1}{\cosh(\theta_{n})}\right)^{D-2}
e^{\tanh(\theta_{n})\,\delta_{I J}\,a_{n}^{\dagger\:I} {\tilde
a}_{n}^{\dagger\:J}} \right]
\left.\left|0\right\rangle\!\right\rangle \label{tva}. \ee Here
$\theta$ denotes the set of transformation parameters. By applying
the operator (\ref{dir}) to the state
$\left.\left|0\right\rangle\!\right\rangle$ one may see that this
state encloses the specific values of the endpoints position, and
furthermore this is not modified by the Bogoliubov transformation.

The thermal creation and annihilation are also transformed according
to

\bea a_{n}^{I}(\theta_{n}) &=& e^{-iG}a_{n}^{I}e^{iG}
=\cosh(\theta_{n})a_{n}^{I} - \sinh(\theta_{n}){\widetilde
a}_{n}^{\dagger \: I}
\\
\widetilde{a}_{n}^{ I}(\theta_{n}) &=& e^{-iG}{\widetilde
a}_{n}^{ I}e^{iG}
= \cosh(\theta_{n}) a_{n}^{I} -  \sinh(\theta_{n}) {\widetilde
a}_{n}^{\dagger \: I}
\eea

As the Bogoliubov transformation is canonical, the thermal
operators obey the same commutation (\ref{alg}).
 These operators annihilate the state
written in $(\ref{tva})$ defining it as the vacuum. By using the
Bogoliubov transformation, the relations
\begin{eqnarray}
a_{n}^{I}(\theta_{n})\left |0(\theta)\right\rangle &=& \widetilde
{a}_{n}^{I}(\theta_{n})\left |0(\theta)\right\rangle = 0,
\end{eqnarray}
give rise the so called thermal state conditions:
\begin{eqnarray}
\left[a_{n}^{I}-
\tanh(\theta_n)\widetilde{a}^{\dagger \: I}_n\right]
\left|0(\theta)\right\rangle
&=&0,
\label{cond1}
\\
\left[\widetilde{a}_{n}^{I}-
\tanh(\theta_n){a}^{\dagger\: I}_{n}\right]
\left|0(\theta)\right\rangle
&=&0,\label{cond2}
\end{eqnarray}

Then, the physical open string Fock space is constructed by applying the
thermal creation operators to the vacuum $(\ref{tva})$
 that may consistently be identified with the D-brane state. In fact,
 if the modes of the open string attached to the D-brane are created from this state,
 precisely in absence of such string excitations, the D-brane on its own must be associated to the fundamental state.

\vspace{0.4cm}

Finally, the thermal open string vacuum is completely defined by
minimizing the free energy
\begin{equation}
F=U-\frac{1}{\beta }S \label{f}
\end{equation}
with respect to the transformation's parameters $\theta$`s \cite{ume2}. Here $U$
is given by computing the matrix elements of the open string
Hamiltonian in the thermal vacuum and $S$ is the expectation value
of the entropy operator $K \equiv -\sum_{n=1} N_{n} \ln N_{n}$ in
this state.  The number operator is defined by
\begin{equation}
N_{n}= a_{n}^{\dagger \: I}  a^J_{n } \,\delta_{I J}\,\,\,, \label{hs}
\end{equation}
  whose expectation value is proportional to $\sinh^2
\theta_n$, therefore we get
\begin{eqnarray}
K &=& -\sum_{n=1} \bigg\{ a_{n}^{\dagger \: I} a^J_{n } \,\delta_{I J}\,\ln \left(
\sinh^{2}\left(\theta _{n}\right)\right) - a^I_{n} a_{n}^{\dagger \: J
} \,\delta_{I J}\, \ln \left( \cosh^{2}\left(\theta _{n}\right)\right)\bigg\}.
\label{k}
\end{eqnarray}
Therefore, the solution for the angular parameters $\theta_n$ is
given by the Bose-Einstein distribution: \be\label{bose} \sinh^2
\theta_n = (e^{\b E_n } -1)^{-1 } .\ee

Notice that, although expressed in the open sector, the thermal
state conditions (\ref{cond1}, \ref{cond2}) together with the thermal
equilibrium requirement (in order to fix the free parameters
$\theta_n$) determine the state of the system so as a D-brane state
is built in the closed channel. So in this sense, we could see
(\ref{cond1}, \ref{cond2}) as a sort of boundary state condition in the
open channel. In Section 5 we will construct a geometrical
approach for boundary states in the open channel, where this
interpretation arises explicitly.

It is easy to see that thermal states are not
eigenstates of the original Hamiltonian but they are
eigenstates of the combination:
\begin{equation}\label{hamiltonian}
{\widehat H} = H -{\widetilde H},
\label{hath}
\end{equation}
in such a way that ${\widehat H}$ plays the r\^{o}le of the
Hamiltonian, generating  temporal translation in the thermal
Fock space.
Let us point out
that the physical variables are described by the non-tilde
operators. This is then the Hamiltonian which governs the dynamical evolution of the D-brane and its excitations, which correspond to
an open string attached to it.

The state (\ref{tva}) describes a condensate of entangled open string modes localized on the D-brane surface. Since this is a coherent state, it
constitutes a macroscopical object (see reference \cite{ume1}) which may be identified with the D-brane. So we conclude this part
 by emphasizing that (\ref{tva}) describes the microscopical structure of the D-brane in terms of open string modes.

\textbf{Analytical Continuation:} Let us remark that despite D-brane
states have been constructed here at arbitrary finite temperature,
the parameter $\b$ can be analytically continued to the complex
plane and in particular to purely imaginary values, $\b= i\lambda\;
, \; \lambda \in \R$ (see ref. \cite{ademir}) in order to describe
states without temperature. In this context however, the quantity
$\tau$ should be not interpreted as a time evolution parameter in
order to describe stationary states, but, as clarified in the
construction of Section 5, it could be seen as sort of ``time delay"
between the physical system and its copy.

\subsection{D-branes States from the current Closed String Description}

We found a way to represent a D-brane as an state in the open channel and we do not need
to transform the problem by going to the closed representation; however
by using the world sheet duality \cite{boundarystates}, one may verify 
 that the thermal vacuum
(\ref{tva}) consistently \emph{corresponds} to a current boundary state in the closed channel. 

Let us first motivate our proposal in the context of this duality. In fact,
  the interaction between two D-branes\footnote{Or, according to
the case analyzed in this paper, the self-interaction of only one
D-brane.} is given by the vacuum fluctuations of an open string
ending on them and propagating in a loop with periodic Euclidean
time $t \in [0,\b]$ (the Casimir effect). Graphically, the topology
of this open string world-sheet is a cylinder ending on the two
branes. Since the theory is conformally invariant, one can find a
conformal transformation such that the world-sheet coordinates are
exchanged and the cylinder corresponds to the tree diagram of a
boundary state of closed string being created in one of the
D-branes, propagated for a while, and annihilated on the other
brane. These boundary states are identified with the D-branes states
in the closed string channel. The crucial observation of our work is
that one may avoid the world sheet transformation in this algorithm,
and to recognize the D-brane states in the open sector. The second
remarkable point in this observation is that, roughly speaking, the one-loop cylinder
diagram in fact characterizes a thermal state by virtue of the
Euclidean period. Let us now show this in detail:

\vspace{0.3cm}

Consider a particular
initial configuration of a closed string,
\be\label{Nt1}
\partial_{t}X^a|_{t=0 }=0 \;\;\; a=0,1,..., p
\ee
\be\label{Dt1}
 X^i|_{t=0 }= x^i \;\;\; i=p+1,...25\; .
\ee
 Because the operators $\partial_{t}X^a \,,\, X^i $ commutate among themselves at the same time,
 a specific configuration of these constitutes a definite \emph{state} in the quantum Hilbert space of closed string.
  This is called a boundary state which is interpreted as the Dp-brane state in itself. It may be expressed as a coherent
   state of closed string modes and its form is remarkably similar to (\ref{tva}) \cite{boundarystates}.

The transition amplitude from this state, correspondent to the initial configuration (\ref{Nt1},\ref{Dt1}) and denoted by
 $| B_p (t=0) \rangle $, into a final one $| B_p (t=-i\pi) \rangle$ (defined by conditions similar to (\ref{Nt1},\ref{Dt1}))
  through an imaginary time interval $-i\pi$, is given by
\be
\langle B_p (t=-i\pi) |  e^{- \pi
 \, H_c } |B_p (t=0) \rangle = \int_\Gamma D X e^{-S_{W_c} [X]}
\ee where $H_c$ is the closed string hamiltonian. This may be represented as a sum over histories as expressed in the right hand side of this identity,
 where $W_c \sim S^{1}_{\b}\times [0, \pi]$ is the closed world-sheet topology whose boundary are two circles that we denote by $S^{1}_{\b,\pm}$.
Then $\Gamma$ represents the set of histories of one closed string ending on these two circles whose states are fixed by configurations as
 (\ref{Nt1},\ref{Dt1}).

 Then, by considering the world sheet transformation $\sigma, t \to t,
\sigma$, $W_c$ transforms into $W_o$, and the above sum over histories coincides with (\ref{path}).
 Finally, one straightforwardly obtains the state (\ref{tva}) according to the procedure previously shown, which 
 encodes the information of the boundary state. In fact, this turns out clear that the state $\left |0(\beta)\right\rangle$,
 so as the boundary state $| B_p  \rangle $, may both be used to calculate the significant
 observables/amplitudes in the respective representations, and they are corresponding in the proper
 sense \footnote{Transition amplitudes between different D-brane states will be considered
 in a forthcoming paper.}. If one inserts any operator $Q$ as in eq. (\ref{tr1}), 
the quantity $\left\langle 0 (\b ) \left| Q
\right| 0 (\b )\right\rangle$,
 must be identified with the amplitude
\be
\,\, A_{B_p}[Q_c] \, = \,\int_{\Gamma[B_p]} D X \,\, Q_{c}\, \,e^{-S_{W_c} [X]} \,\, ,\label{QC}
\ee computed in the closed channel. This integral is a sum over all worldsheet embeddings subject
 to the boundary conditions (\ref{Nt1},\ref{Dt1}) as discussed above. In particular, if we take $Q$ to be  
 a product of operators $\{X(\sigma_i, t_i), i=1,...n\}$ valued on a collection of $n$ different
 world-sheet points, we get the $n$-point thermal Green function of the open string \footnote{The propagators in the thermal vacuum}. On the other hand, $Q_{c}$ corresponds to the same object in the closed channel, the $n$-point
correlation function for $n$ points of the closed string, which are defined by exchange of the
world-sheet coordinates: $(\sigma_i , t_i)_{open} \to (t_i , \sigma_i)_{closed}$, according to discussion above. This prescription may be extended to consider products of different derivatives of these operators.
 This completes the argument on our inicial statement.

\section{Geometric formulation and boundary states in the open string representation.}

The goal of this section is to show that D-brane states may be constructed as boundary states
even in the open sector in an appropriate sense. In other words, the
boundary state conditions whose solution is $\left|B_{open}
\right\rangle$ may be imposed also in the open string channel as
conditions on states rather than operators in a way similar to the Gupta-Bleuler standard procedure.
 In fact, these conditions consist in a fixing of non-physical variables in terms of physical ones\footnote{which may alternatively be interpreted as a selection of physical states, that in this case shall describe the D-brane and its excitation (open) modes at finite temperature.} (on their respective spatial boundaries), which is analogous to a gauge fixing.  So in this study, we rigorously refer to boundary state in the context of open string in this precise sense; however, we would like to emphasize here that despite these states may be fixed initially\footnote{Then they are preserved by the evolution of the system, generated by the total Hamiltonian (Eq (\ref{hamiltonian}))} and, as argued before they carry the same information as the closed string boundary states, it should be clearly differentiated from the concept of ``boundary state in the open string channel'' properly introduced in Ref. \cite{matsuo}, where such states may actually describe the emission and absorption of the 
(open) strings by D-branes as the usual boundary state does for the closed strings.

To do this we introduce a purely geometrical approach which only requires the duplication
 structure of TFD and where the thermodynamic concepts may be ignored. The dynamical information
  of the system is sufficient to determine these states.
 This technique is interesting in itself because it seems to be applicable to other situations with
  boundary conditions (e.g. \cite{ademir}), it is similar to TFD but incorporates some new ingredients
   related with the dynamical properties of the system and with its geometry.

Note that the main idea underlying the TFD approach to this problem (emphasized in this new construction)
 is the possibility of describing the contact of one open string with a D-brane by effectively substituting
  such object, whose dynamics and degrees of freedom are unknown in principle, by another fictitious string.
   In fact, it seems to be natural to think that the effective degrees of freedom of the brane, which are
    activated by the energy-momentum exchange with the physical string, be in correspondence with a single-string
     degrees of freedom. This is what we call the fictitious string or ``hole" (as used in the TFD literature).

In geometric terms, we may represent a string ending on the D-brane
surface, while the fictitious string lives on the other side \cite{laf} and
  ends on the \emph{same} brane as required
by the TFD duplication rules. The group of invariance of the string
attached to a Dp-brane is $G_p \equiv SO(1, p ) \times SO( D-p)$ and
another equal symmetry ${\tilde G_p}$ should be attributed to the
fictitious string variables.

The boundary conditions to quantize the open string are
\be\label{dir2}
 X^i|_{\s=0,\pi  }= x^i \;\;\; i=p+1,...25
\ee \be\label{neu2}
\partial_\s X^a|_{\s=0,\pi  }=0 \;\;\; a=0,1,..., p
\ee and they must be the same for $\X$, in order to have a copy of
the original system as required by TFD:  \be\label{tildedir2}
 \X^i|_{{\tilde \s}=0,\pi  }=  x^i \;\;\; i=p+1,...25
\ee \be\label{tildeneu2}
\partial_\s \X^a|_{{\tilde\s}=0,\pi  }=0 \;\;\; a=0,1,..., p
\ee
which defines another string in contact with a Dp-brane in the same position $x^i $. Once more we
 assume the Hilbert space ${\cal H}_{o}\otimes\widetilde{{\cal H}}_{o}$. By using a part
of the symmetry ${\tilde G_p}$ we may translate its endpoint along the Dp-brane hyperplane to coincide with those of
the physical string. So we may see this procedure as a sort of gauge fixing which
according to the Gupta-Bleuler prescription,
this shall to be imposed on states as follows:
 \be\label{bound-dir-open} {\X}^\mu ({\tilde\s}, {\tilde t})|_{{\tilde\s}=0,\pi } - X^\mu (\s,t)|_{\s=0,\pi } \;\;\left|B_{open}
\right\rangle = 0 \,\,\, ,\ee
whose components $\mu=i=p+1,...25$ are trivially satisfied due to
(\ref{dir2}) and (\ref{tildedir2}), furthermore this manifestly implies that the respective
 RHS of these two conditions must coincide. On the other hand, in order to
 ensure the smooth gluing of both open strings in their respective endpoints,
 we shall require the continuity of the first derivative with respect to the respective strings parameters in their boundaries. This
 condition is:
 \be\label{bound-neu-open} \partial_{\tilde\s} {\X}^\mu ({\tilde\s}, {\tilde t})|_{{\tilde\s}=0,\pi } -
  \partial_\s X^\mu (\s,t)|_{\s=0,\pi } \;\;\left|B_{open}
\right\rangle = 0 \,\,\, ,\ee
whose components $\mu=a=0,...., p+1$ trivially anihilate all the states of the Hilbert space
by virtue of
(\ref{neu2}) and (\ref{tildeneu2}). Then (\ref{bound-dir-open}) and (\ref{bound-neu-open}) constitute $D$ non trivial
 conditions on the D-brane states $\left|B_{open}
\right\rangle$ ($D-2$, considering only the physical components in the light cone gauge).

Let us remark that
the tilde variables were fixed in terms of the non-tilde ones by these conditions, which clearly
break the symmetry $G_p \times {\tilde G_p} $ into the original group $G_p$.

The map between the tilde and non-tilde operators is
defined by the following tilde (or dual)
conjugation rules \cite{kha5}:
\begin{eqnarray}
( X Y)\widetilde{}
&=&\widetilde{X}\widetilde{Y}, \nonumber \\
(c\, X + Y)\widetilde{} &=&c^{\ast
}\,\widetilde{X}_{i}+\widetilde{Y}_{j}, \nonumber \\
(X^{\dagger })\widetilde{}
&=&(\widetilde{X})^{\dagger }, \nonumber \\
(\widetilde{X})\widetilde{} &=& X,  \nonumber \\
\lbrack \widetilde{X},Y] &=&0.  \label{til}
\end{eqnarray}

Considering the solutions (\ref{stringsoldir}) and (\ref{stringsolneu}) in the light cone gauge,
 one may use these rules to construct the fictitious copy of the string $\X^I$ defined on an
  independent (\cite{laf}) world sheet manifold whose coordinates are ${\tilde t},{\tilde \s}$, and we get:
\be\label{stringsoldir-tilde-invert}  \X^i ({\tilde t}, {\tilde \s})= x^i_{-}
\;(\pi-{\tilde\s})/\pi + x^i_{+ } \; ({\tilde\s}/\pi) - \sqrt{2\a'}
\sum_{n \neq 0}\left(\frac{1}{n}\; {\tilde \a}_n^i \;e^{in {\tilde t}}\; \sin(n {\tilde\s})\right) ,\ee
\be\label{stringsolneu-tilde-invert} \X^a({\tilde t}, {\tilde\s})= {\tilde x}^a + 2\a' {\tilde p}^a {\tilde t} - 2i \a'
\sum_{n \neq 0} \left(\frac{1}{n}\; {\tilde \a}_n^a \; e^{in {\tilde t}} \;\cos(n {\tilde\s})\right) .\ee
If we finally redefine the mode number $n\to -n$ \footnote{Notice that this change is related to the
 structure of the Hamiltonian (\ref{hamiltonian}) for the duplicated system.} in each term of this expression, the solution reads:
\be\label{stringsoldir-tilde}  \X^i ({\tilde t}, {\tilde \s})= x^i_{-}
\;(\pi-{\tilde\s})/\pi + x^i_{+ } \; ({\tilde\s}/\pi) - \sqrt{2\a'}
\sum_{n \neq 0}\left(\frac{1}{n}\; {\tilde \a}_{-n}^i \;e^{-in {\tilde t}}\; \sin(n {\tilde\s})\right) ,\ee
\be\label{stringsolneu-tilde} \X^a({\tilde t}, {\tilde\s})= {\tilde x}^a + 2\a' {\tilde p}^a {\tilde t} + 2i \a'
\sum_{n \neq 0} \left(\frac{1}{n}\; {\tilde \a}_{-n}^a \; e^{-in {\tilde t}} \;\cos(n {\tilde\s})\right) .\ee

Although the time evolution of both strings is respectively given by
the independent time parameters $t$ and $\T$, they both shall
parameterize the same time direction (in the target) in order to
preserve the gauge choice; then
 in particular, we may define them up to a general shift $ \T \equiv t + \t$. In the light cone frame both time
  parameters are given by the coordinate $X^+$ and they only can be related up to a additive number. Below, we
   will discuss on the important meaning of this time delay between both strings.

An appropriate initial choice of the non-physical/physical gluing of
the variables in the boundary of the system
 according to the conditions  (\ref{bound-dir-open}) and (\ref{bound-neu-open}) shall in fact determine
  boundary states, then these states shall be built such that these conditions be satisfied for all $t$.
   By requiring this in the expansion in modes of (\ref{bound-dir-open}, \ref{bound-neu-open}), results:
\be\label{B1} \A^{I}_n - e^{i\,n\, \t}\;a^{\dagger\:I}_n
\;\;\;\left|B_{open} \right\rangle = 0 , \ee
\be \label{B2}x^a - {\tilde x}^a + 2\a' {\tilde p}^a \t \;\;\left|B_{open} \right\rangle = 0 , \ee
\be \label{B3} {\tilde p}^a - p^a \;\;\left|B_{open} \right\rangle =
0 ,\ee
where we have used (\ref{redef}) to define the canonical creation/anihilation operators.
Then the solution
of these equations may be expressed as: \be\label{boundary} \left|B_{open}
\right\rangle = N'
\delta(x^a - {\tilde x}^a + 2\a' {\tilde p}^a \t) \delta({\tilde p}^a - p^a)\prod_{I, n>0} \,\,
 e^{q_n a^{\dagger\:I}_{n} \A^{\dagger\:I}_{n} } \left.\left|0\right\rangle\!\right\rangle \, ,\ee
where $q_n =
e^{i\,n\,\t}$, $N' \equiv \; N \delta(X^i|_{\s=0,\pi} - x^i_{\pm})\delta(\X^i|_{{\tilde\s}=0,\pi} - x^i_{\pm})$
 and $N$ is the normalization constant\footnote{Note that the centers of mass of both strings follow the same trajectory but
the position of one of them is forwarded by $2\a' {\tilde p}^a \t$ with respect to the other one.}.

\vspace{0.5cm}

Now we remarkably observe that the time shift $\t$ may be an arbitrary complex number.
In particular if this is taken to be a purely imaginary number $\t \equiv -i \b/2$, one may
 define the physical time $t$ parameter to be real and in this case, the evolution of
  fictitious system will be parameterized precisely by the real part of ${\tilde t}$ (where $\t \equiv Im ({\tilde t}) $).

If $\t \equiv -i \b/2$, and if one assumes reality conditions also for the fictitious variables, equation (\ref{B2}) splits into the equations:
\be \label{B2x}x^a - {\tilde x}^a  \;\;\left|B_{open} \right\rangle = 0 \, ,\ee
\be \label{B2p} {\tilde p}^a \;\;\left|B_{open} \right\rangle = 0\, .\ee

Thus the solution
 of (\ref{B1}\,,\,\ref{B3}\,,\,\ref{B2x}\,,\,\ref{B2p})
  may be written as: \be\label{boundarytermico} \left|B_{open}
\right\rangle = N'
\delta(x^a - {\tilde x}^a) \delta({\tilde p}^a) \delta(p^a)\prod_{I, n>0} \,\,
 e^{q_n a^{\dagger\:I}_{n} \A^{\dagger\:I}_{n}} \left.\left|0\right\rangle\!\right\rangle \, .\ee

In this case $q_n = e^{- n\,\b/2}$. Defining
$ \theta_n = \tanh^{-1} q_n( \t)$, the string modes occupation
number is given by $N_n = \sinh^2 \theta_n $, which agrees with a
Bose-Einstein distribution of string modes at the temperature
$\b^{-1}$, and this state remarkably coincides with (\ref{tva}) found in Section 2. Therefore, we showed here that the effect
\emph{produced} by the brane on the open string at finite temperature is equivalent to the effect due to another string,
 joined to the first one through their boundaries, but forwarded by an imaginary time interval (which in this formalism need not
  to be compactified to the circle as usual \footnote{In particular in the imaginary time approach \cite{mat1} \cite{adas}}).
   Furthermore this time delay may be interpreted in terms of the temperature of the brane.

As pointed out in Section 2, one may consider the analytical continuation of the parameter ${\tilde \t}$ to
 take real values and the boundary states are given by (\ref{boundary}).
  Although this does not arise from the traditional TFD construction, we may observe here that this number
   should not be interpreted as a time evolution parameter (as often believed) but as a relative delay.

   Finally, it is interesting to remark that this may be seen as the geometrical representation of the
    postulate of non-physicalness of the tilde system, since the ``world-manifold" associated with it,
     is causally disconnected of the parallel world-manifold of the real system due to the imaginary
      time delay between them. However there is quantum entanglement between them which give place to
       thermal states. In contrast if $ \t\in \R $, in principle both systems may superpose and
        to cause interference among them. This may be describing dynamical (out of the thermal equilibrium)
         effects of the brane. We furthermore believe that this also open the possibility of studying interactions at finite temperature \cite{Gaume}.

 These are some of the new remarks and perspectives emphasized by this formulation and shall be studied in detail
  elsewhere\footnote{We shall to cite that the construction presented here and the geometric description
   of thermal closed string given in \cite{gluing-thorus} have some similar aspects. }.

\section{Concluding Remarks and Outlook}

In this work we obtained the open string representation of a bosonic
Dp-brane state and its generalization at finite temperature. A
remarkable strength of this description is that the Dp-branes are
emphasized as vacuum states of a open string Fock space. On the
other hand, this approach handles a model for Dp-branes where its
macroscopical nature is manifest \cite{ume1}. Notice in particular
that if we consider an ensemble of open strings attached to the same
Dp-brane, then in the thermodynamic limit this may be seen as a sort
of \emph{medium} extended on $p$ spacial dimensions, filled with
open string modes.

Let us observe that the configuration discussed
in Section 5 constitutes a composite between two open strings (so
one may say that $\left|B_{open} \right\rangle$ represent composite
states) and notice in addition, that from a strictly topological
point of view, they two form a closed string.
 This observation could help to understand what is the manifest correspondence between both
  open/closed representations of D-brane states in the TFD language and this furthermore
   gives rise to a more deep question: may closed string states be viewed as bound states of open strings?

In this sense, we believe that the open/closed duality could be properly interpreted as a
 correspondence between closed string states and \emph{thermal} (mixed) states of open strings. In fact,
according to the arguments explained in Section 3, if we search for an open string configuration at thermal
 equilibrium, we generically may read it as a path integral formulation of a closed
  string theory. On the other hand, let us remark that this fact suggests that the kinematical structure
 of an open string field theory shall require a $C^{\star}$ algebra \cite{haag,emch}, and indicates how
  the closed string sector could be recovered in such theory.

In future works we will investigate the thermal stability of the
D-brane configurations so as the possibility of describing black
branes using these ideas. According to the references
\cite{bb1,bb2,bb3}, The infrared behavior of theories whose dual
bulk-gravities contain a black brane is governed by hydrodynamics,
and the main observation in this sense is the existence of an
universal value for the ratio of shear viscosity to entropy density
\cite{universal} which should be investigated in the context of an
appropriate microscopical model.

 Finally, the connection of our results with the closed string description of
  thermal D-branes \cite{IVV,AGV2,AEG,AGV3,AGV4,AGV5} should be clarified.

\section{Acknowledgements}

The author would like to thank to A. Gadelha,  N.
Grandi and  G. Silva for fruitful conversations on the subject of
this paper. J. A. Helayel-Neto, R. Scherer Santos and N. Quiroz Perez, are specially aknowledged for useful comments and observations.
This work was supported by CONICET.

 \end{document}